\documentclass[12pt,superscriptaddress,english,prd,nofootinbib,notitlepage]{revtex4-1}
 \pdfoutput=1
\usepackage[footnotesize,singlelinecheck=off,justification=raggedright]{caption}
\usepackage[english]{babel}
\usepackage[latin1]{inputenc}
\usepackage[colorlinks=true]{hyperref}
\usepackage{url}
\usepackage{graphicx,soul}
\usepackage[dvipsnames]{xcolor}
\usepackage{subfig}
\usepackage{latexsym,xfrac}
\usepackage{amsmath}
\usepackage{amsfonts,dsfont}
\usepackage{amssymb,slashed}
\usepackage{bbold}
\usepackage{verbatim,epigraph}
\usepackage{tikz-cd} 
\usepackage{enumitem}

\usepackage{float}
\usepackage{dcolumn}
\usepackage{bm}
\usepackage[format=plain,justification=justified]{caption}

\usepackage{ulem}

\begin{document}

\title{Entanglement in a Maxwell Theory
       coupled to a non-relativistic particle}

\author{Filiberto Ares}
\email{fares@iip.ufrn.br}
\affiliation{International Institute of Physics, Universidade Federal do Rio Grande do Norte, 59078-970, Natal, RN, Brazil}

\author{Amilcar R. de Queiroz}
\email{amilcarq@gmail.com}
\affiliation{Instituto de F\'isica, Universidade de Bras\'ilia, 70919-970, Bras\'ilia, DF, Brazil}

\author{Marcia R. Tenser}
\email{marciatenser@gmail.com}
\affiliation{Instituto de F\'isica, Universidade de Bras\'ilia, 70919-970, Bras\'ilia, DF, Brazil}
\affiliation{Instituto de F\'isica, Universidade de S\~ao Paulo, 05315-970, S\~ao Paulo, SP, Brazil}

\begin{abstract}
  We consider electromagnetism in a cylindrical manifold coupled to a
  non-relativistic charged point-particle. Through the relation between 
  this theory and the Landau model on a torus, we study the
  entanglement between the particle and the electromagnetic field.
  In particular, we compute the entanglement entropy in the ground state,
  which is degenerate, obtaining how it varies in the degeneracy
  subspace.
\end{abstract}

\maketitle

\section{Introduction}

Being the characteristic trait of quantum mechanics
\citep{Schrodinger}, entanglement has been revealed as
fundamental in many fields and phenomena ranging
from quantum information \cite{Nielsen, Horodecki}, 
and condensed matter \cite{Amico, CalCarDoy, Laflorencie} 
to black hole physics \cite{Solodukhin}. One of the standard 
quantities employed to characterise entanglement is entanglement 
entropy (EE). In order to define it, we must partition the 
system into two parts such that its Hilbert space is 
the tensor product of the Hilbert space of the partitions. 
Then the EE measures the degree of entanglement 
between the two subsystems \cite{Bennett}.

In field theories, one possibility is to divide 
the real space into several regions. This was originally 
done in a scalar field theory \cite{Sorkin, Srednicki} 
motivated by the Bekenstein-Hawking formula 
for a black hole. Later the analysis of spatial
entanglement was extended to conformal field theories
\cite{Holzhey, CarCal} that the Ryu-Takanayagi formula
\cite{Ryu} connects with gravity \cite{Raamsdonk} via the
holographic principle. Spatial EE in massive field 
theories has also been well studied, see for example
\cite{Casini1, Casini2, Casini3, Casini4, Cardy, Castro-Alvaredo}.
One can also consider the entanglement between other
partitions of the Hilbert space that are not spatial 
ones, as, for example, between right and left moving 
excitations \cite{Pando, Das, Lencses} or between 
winding modes \cite{Prudenziati}. 

When the field theory presents gauge symmetry, 
spatial partitions are subtle: it is not possible to 
make them and still preserve gauge invariance. This 
difficulty is due to the fact that gauge theories contain 
non local degrees of freedom such as Wilson loops. Hence, 
when a spatial partition is made, these loops are 
necessarily broken. So we are left with an arbitrary 
choice (and therefore an ambiguity) of deciding to which 
of the subregions the broken degrees of freedom belong. 
Many aspects of this problem have been addressed 
in the literature since the work \cite{Kabat}. 
It has been discussed in the context of lattice gauge theory giving
rise to different prescriptions for computing EE
\cite{Buividovich, Donnelly, Trivedi, Trivedi2, Aoki, Lin}. 
In the continuum, one possibility is to calculate EE
using the replica trick after extending the Hilbert 
space in a particular way \cite{Gromov, Velytsky,  Donnelly2}. 
Without resorting to the replica trick, some alternative approaches 
have also been considered. For example, in \cite{Nair}, 
EE in 2+1 dimensions was studied employing gauge-invariant
variables. In \cite{Casini5}, the zoo of prescriptions
for computing EE was unified using an algebraic approach, defining
it in terms of a subalgebra of gauge-invariant operators associated
to each subregion. Another algebraic framework based on the 
Gel'fand-Naimark-Segal construction was suggested formerly in 
\cite{Balachandran1, Balachandran2} in order to treat systems 
of identical particles. This method was also applied to analyse 
the ambiguities of EE in systems with gauge symmetries 
\cite{Balachandran3, Balachandran4}. Recently in \cite{casini2019logarithmic}
the authors proposed that a proper measure for spatial entanglement in
a Maxwell theory is mutual information. As one of the motivations to use this quantity
instead of the bare entanglement entropy, they argue that it
resolves the aforementioned ambiguities. This idea had been presented previously
by the same authors in the context of systems with global symmetries in
 \cite{casini2019entanglement}.

The works mentioned in the previous paragraph concern pure gauge 
theories, without coupling to matter. By including matter, one may
not only study spatial entanglement \cite{Aoki2} but also the 
entanglement between the gauge field and the matter sectors. The 
present paper is dedicated to the latter situation. This problem 
has also been investigated recently in \cite{Makarov} where 
the entanglement between a quantum harmonic oscillator and a quantized 
electromagnetic field was analysed. 

Here we consider a non-relativistic particle coupled to an Abelian
Yang-Mills (YM) theory in $1+1$ dimensions with compactified spatial
coordinate, i.e., space-time is a cylinder $\mathds{R}\times S^1$.
In order to compute the EE between the particle and the field,
we map the theory to a quantum mechanical system consisting of a charged
particle moving on a torus with a uniform transverse magnetic flux.
This is the Landau problem \cite{Landau} on a torus \cite{Manton}. 
In fact, as shown in \citep{Rajeev}, the field dynamics of
a pure YM theory defined on a cylinder can be reduced to that of a
free particle moving along the gauge group manifold. This means
that we can reduce the quantum field theory problem to a quantum
mechanical one. For simplicity, here we restrict to the Abelian
case, in which the field theory is mapped to a particle moving 
on a circle \cite{Asorey}. From the field theory point of view, 
the only gauge-invariant observable is the Wilson loop along $S^1$. 
If we were to partition this circle to compute some kind of 
spatial entanglement, then we would break the gauge invariance
and we would need to apply the techniques cited above. The fact 
that there is only one gauge invariant observable in the field 
theory implies that there is a single degree of freedom associated 
to the gauge field in the quantum mechanical theory. Thus if we 
do not consider matter, we have only one degree of freedom 
(a particle moving on a circle) and, therefore, it is impossible 
to make any partitions in this setup. Including a non-relativistic 
particle adds another degree of freedom (the particle moves now on 
a torus) and the possibility of making a partition. Hence the goal 
of this work is to understand the entanglement between the degree 
of freedom associated to the gauge field and that corresponding to 
the non-relativistic particle. 

The paper is organised as follows: in the next section, we show the
equivalence between electromagnetism on a space-time cylinder coupled
to a non-relativistic charged point particle and the Landau problem
on a torus. In section \ref{sec:manton}, we obtain the solution of 
the Schr\"odinger equation of the latter, finding that the ground 
state is degenerate. In section \ref{sec:entanglemententropy},
we study the entanglement entropy in this degeneracy subspace. This
is equivalent to measuring the entanglement between the particle and
the electromagnetic field in the ground state of the field theory.
In particular, we perform both an analytical and a numerical analysis of
this quantity. We find that the reduced density matrix in the degeneracy 
subspace can be approximated by that of a two-level system. This observation 
allows us to obtain an analytical expression for the entanglement entropy 
in this subspace whose accuracy is checked numerically. We also study the 
entanglement entropy in the state that is invariant under the symmetry 
transformation associated to the degeneracy subspace. Finally, in section 
\ref{sec:conclusions}, we present the conclusions and outlooks. We include 
an Appendix \ref{appendix} where we compute the Green's function of the 
electric potential in the Maxwell theory on the space-time cylinder.

\newpage 

\section{Electromagnetism on a Space-time Cylinder and the Landau model on
a torus}
\label{sec:ym}

In this section, we introduce the model to be discussed later and fix
the notation. The model is electromagnetism on a space-time cylinder
coupled to a non-relativistic charged point particle. We review its
relation with the Landau model on a torus, that is, a charged particle moving
on a torus subject to a transverse magnetic field. 

\subsection{Electromagnetism Coupled to a Charged Point Particle}
\label{sec:electro}

We consider a space-time cylinder with coordinates 
$\mathbf{s} =(s^0,s^1) \equiv (t,s)$, where $t\in\mathds{R}$ 
and $s\in [0,2\pi R)$. If $\tau$ is the proper time of the point particle, with 
electric charge $q$ and mass $m$, then its classical 
trajectory can be parametrized as ${\bf r}(\tau)=(r^0(\tau), 
r^1(\tau))\equiv(t(\tau), r(\tau))$. The electromagnetic 
field is described by $F_{\mu \nu} = \partial_\mu A_\nu - 
\partial_\nu A_\mu$, $\mu,\nu=0,1$. The electric field is $E=F^{01}$. 
The field equations are $\partial_\nu F^{\mu\nu} = J^\mu$, 
where\footnote{$\delta^{(2)}(\mathbf{s}-\mathbf{r}(\tau))=
\delta(s^0-r^0(\tau))~\delta(s^1-r^1(\tau))$.}
\begin{equation}
 J^\mu(t,s) = q\int {\rm d}\tau~\delta^{(2)}\big(\mathbf{s}-
 \mathbf{r}(\tau)\big)~\frac{{\rm d} r^\mu(\tau)}{{\rm d}\tau}\,.
\end{equation}
If we align the proper time $\tau$ with the time coordinate 
of the particle, so that $t=\tau$, then we have $J^0=q~\delta\big(s-r(t)\big)
\equiv \rho$ and $J^1=q~ \delta\big(s-r(t)\big)\dot{r} \equiv j$. 
Therefore, in local coordinates, the field equations read
\begin{align}
\partial_{s} E &
=-\partial_{s} \partial_{t} A_1 
+ \partial_{s}^2 A_0 = \rho\,, \label{Gauss-Law-1} \\
\partial_{t} E &=-\partial_{t}^2 A_1 + \partial_{t}
\partial_{s} A_0   = - j \label{Ampere-Law-2}\,.
\end{align}
For simplicity, we consider a non-relativistic charged point particle. 
Its equation of motion reads\footnote{$m\ddot{r}^\mu=q F^{\mu\nu} \dot{r}_\nu$.} 
$m\ddot{r}=q E$. From now on, we set $m=1$.

The above equations of motion are obtained from the Lagrangian
\begin{equation}\label{lagrangian-ym}
 L=L_{\rm M}+L_{\rm EM} = \frac{ \dot r^2}{2}
 + \int {\rm d}{s}~\left(\frac{\left(\partial_{t} A_1 - 
 \partial_{s} A_0\right)^2 }{2} + A_0 \rho + A_1 j\right) .
\end{equation}
In the Coulomb gauge, $\partial_{s} A_1 = 0$, the Gauss law (\ref{Gauss-Law-1}) 
becomes $\partial_{s}^2 A_0 = \rho$. Therefore, $A_0$ is not dynamical
and its only role is enforcing the Gauss law. This constraint can be readily solved as
\begin{equation}
\label{A0-sol-w-Green-1}
 A_0(s) = \int {\rm d}s'~ G(s,s')~\rho(s')\,,  
\end{equation}
where $G(s,s')$ is the Green's function of the operator $\partial_{s}^2$. 
Generically, finding this Green's function depends on the boundary conditions.
We will take them to be periodic, such that $A_0(s+2\pi R)=A_0(s)$, 
$\rho(s+2\pi R)=\rho(s)$, and, therefore, $G(s+2\pi R, s')=G(s, s'+2\pi R)
=G(s, s')$. In Appendix \ref{appendix}, we describe in detail the calculation of 
$G(s, s')$. We find
\begin{equation}\label{Green}
G(s, s')=
-\frac{([s]-[s'])^2}{4\pi R}
+\frac{|[s]-[s']|}{2}
-\frac{\pi R}{6},
\end{equation}
where the notation $[\bullet]$ stands for 
$[\bullet]\equiv \bullet\bmod 2\pi R$, and
$s,s'\in \mathds{R}$.

In the Coulomb gauge, the field $A_1(t,s)$ does not depend on the 
spatial coordinate $s$, i.e., $A_1(t,s)=a(t)$. Moreover, gauge invariance 
implies that $a(t)$ is valued on a circle of length $\dfrac{1}{eR}$. In fact, 
consider a gauge transformation $g={\rm e}^{{\rm i}e\Lambda(t,s)}$, where $e$ denotes
the elementary electric charge, that winds around the spatial dimension. 
In order to be a single-valued transformation, $\Lambda(t,s)$ must satisfy
\begin{equation}
\Lambda(t,s=2\pi R)=\Lambda(t,s=0) + \frac{2\pi n}{e}\,, 
\quad\mbox{for some } n \in \mathds{Z}.
\end{equation}
A possible solution is $\Lambda(t,s)=\dfrac{n s}{e R}$. In this case, 
the gauge field transforms as
\begin{equation}
\label{periodicitycondition}
A_1(t,s)\mapsto A_1(t,s) + \partial_s \Lambda(t,s) = 
A_1(t,s)+ \frac{n}{e R}\,.
\end{equation}
The equivalence of configurations of the field 
related by gauge transformations implies that 
we can restrict $a(t)$ to $0\leq a(t)< \dfrac{1}{e R}$.

Now going back to the Lagrangian, the Coulomb gauge allows 
us to rewrite $L_{\rm EM}$ as
\begin{equation}
L_{\rm EM} =\int_0^{2\pi R} {\rm d}s~\bigg(\frac{\dot{a}^2}{2} 
+ \frac{A_0 \rho}{2} + j a \bigg)\,,
\end{equation}
where we have used that $\partial_s^2 A_0 = \rho$ and 
the periodicity of $A_0(s)$.
Applying (\ref{A0-sol-w-Green-1}), we then obtain
\begin{equation}
 L_{\rm EM}=\int {\rm d}s \left(\frac{\dot a^2}{2}+j a \right) 
 + \frac{1}{2}\int {\rm d}s {\rm d}s' \rho(s) G(s,s') \rho(s')\,.
\end{equation}
Taking into account that, according to (\ref{Green}),
$G(r(t), r(t))=-\pi R/6$, the Lagrangian (\ref{lagrangian-ym}) 
can be written as
\begin{equation}
 L= \frac{\dot r^2}{2} + \frac{\dot a^2}{2} 2\pi R 
 + q \dot r a - \frac{q^2\pi R}{12}\,.
\end{equation}
The term $-q^2\pi R/12$ can be dropped since it does 
not affect the equations of motion.
After completing the square, we finally arrive at
\begin{equation}\label{lagrangian}
 L= \frac{\dot a^2}{2} 2\pi R + \frac{1}{2}\left(\dot r 
 + q a \right)^2-\frac{q^2}{2} a^2\,.
\end{equation}

From now on, we define the elementary electric charge as 
$e=2\pi$ and write the charge of the particle as $q=-e\theta
=-2\pi\theta$. We also choose the specific value 
$R=\dfrac{1}{2\pi}$, such that the spatial direction 
has unit length. Observe that, under these considerations,
the gauge field $a(t)$ is also valued on a circle of length one. 
Moreover, $a$ may be replaced by $x$ and the position of 
the particle $r$ by $y$. Therefore, $x\in [0,1)$, $y\in [0,1)$ and the Hamiltonian 
corresponding to the Lagrangian (\ref{lagrangian}) reads
\begin{equation}
\label{Hamiltonian-Landau-on-Torus-1}
 H=\frac{p_x^2}{2} + \frac{1}{2}\left(p_y+2 \pi \theta x \right)^2\,,
\end{equation}
where
\begin{equation}
 p_y=\dot y-2\pi\theta x \qquad {\rm and} \qquad p_x=\dot x\,.
\end{equation}

Since we are interested in studying the entanglement between the 
electromagnetic and the matter sector, in the following sections 
we will consider the quantum version of the Hamiltonian 
(\ref{Hamiltonian-Landau-on-Torus-1}). It can be straightforwardly 
found through canonical quantization, i.e., by promoting $x,\,y$ and $p_x,\,p_y$ 
to operators acting on $L^2([0, 1)\times[0, 1))$ that satisfy the 
canonical commutation relations $[x, p_x]=[y, p_y]={\rm i}$ and 
$[x, p_y]=[y, p_x]=0$.

\subsection{Landau Problem on a Torus}
\label{sec:manton}

The Hamiltonian (\ref{Hamiltonian-Landau-on-Torus-1}) is 
also obtainable from the problem of a particle of charge 
1 moving on a torus with local coordinates $0\leq x,y<1$ 
in presence of a constant magnetic field $B_z=2\pi\theta$
in the transverse direction. This is the famous Landau problem 
on a torus. In Appendix B of \cite{Manton} this model is studied 
in detail for the case $\theta=1$ (see also \cite{Asorey, Esteve1, Onofri}). 
Note that the Hamiltonian (\ref{Hamiltonian-Landau-on-Torus-1}) is 
written in the gauge $\mathcal{A}_y=2\pi\theta x$ and $\mathcal{A}_x=0$, 
so that $B_z=\partial_x \mathcal{A}_y-\partial_y \mathcal{A}_x=2\pi\theta$ 
(we denote the gauge field as $\mathcal{A}$ to avoid confusion with section 
\ref{sec:electro}). Moreover, we can define the momenta
\begin{equation}\label{trans_pi}
 \pi_x=p_x + \mathcal{A}_x=p_x, \qquad {\rm and} 
 \qquad \pi_y=p_y + \mathcal{A}_y= p_y + 2\pi \theta x\,.
\end{equation}

There is a second set of translation operators that 
commute with the above momenta and, therefore, with 
the Hamiltonian (\ref{Hamiltonian-Landau-on-Torus-1}). 
These translations are generated by
\begin{equation}\label{trans_v}
v_x=p_x+2\pi\theta y, \qquad {\rm and} \qquad v_y=p_y\,.
\end{equation}

Although all the operators in (\ref{trans_pi}) and (\ref{trans_v}) 
formally commute with the Hamiltonian (\ref{Hamiltonian-Landau-on-Torus-1}) 
as differential operators, the system presents an anomaly and not all 
the translations generated by them are symmetries of the theory due 
to the boundary conditions \cite{Esteve1, Esteve2}. The Hamiltonian 
(\ref{Hamiltonian-Landau-on-Torus-1}) is self-adjoint if the 
wave functions satisfy
\begin{equation}\label{boundaryconditiononx}
 \psi(0,y;t)={\rm e}^{2\pi {\rm i} \theta y } ~ \psi(1,y;t)\,,
 \quad
 \psi(x,0;t)=\psi(x,1;t\,).
\end{equation}
Similar conditions apply to the first derivatives of the wave 
function. These boundary conditions define the domain of the 
Hamiltonian.

That fact implies that translational invariance is broken to the discrete
cyclic subgroup $\mathds{Z}_\theta\times\mathds{Z}_\theta$, with 
$\theta\in\mathds{Z}$ (see \cite{Onofri} for a comprehensive 
discussion). The infinitesimal translations generated 
by $v_x$ and $v_y$ do not respect the boundary conditions; 
that is, their images are states which in general do not fulfil 
(\ref{boundaryconditiononx}) and, therefore, they are not in the 
domain of the Hamiltonian (\ref{Hamiltonian-Landau-on-Torus-1}).
As shown in \cite{Esteve1}, an operator is anomalous 
if it does not keep invariant the domain of the Hamiltonian. 
Only the discrete translations
\begin{equation}\label{discrete_translations}
V_x(l)={\rm e}^{{\rm i}\frac{l}{\theta} v_x},
\quad 
V_y(l)={\rm e}^{{\rm i}\frac{l}{\theta} v_y},
\quad l=1,\dots,\theta,\,\, \theta\in\mathds{Z},
\end{equation}
map the domain of the Hamiltonian (\ref{Hamiltonian-Landau-on-Torus-1}) 
into itself. As a consequence, they are the only translations that are 
actual symmetries of the theory.
 
First of all, let us see that $\theta$ has to be an integer. 
Recall that the length of the torus in each 
direction is $1$. Hence a full rotation around each direction 
of the torus is performed by 
\begin{equation}
 V_x\equiv V_x(\theta) = {\rm e}^{{\rm i} v_x} \qquad {\rm and} 
 \qquad V_y\equiv V_y(\theta)={\rm e}^{{\rm i} v_y}\,.
\end{equation}
Now, if one starts with a wave function satisfying 
(\ref{boundaryconditiononx}) at the point $(x,y)\equiv (0,0)$ 
and transports it to $(1,1)$, there are two possible paths,
\begin{align}
 (0,0)\to(1,0)\to(1,1)\,, \qquad V_y V_x \psi(0,0) &= V_y \psi(1,0)=\psi(1,1)\,, \\
 (0,0)\to(0,1)\to(1,1)\,, \qquad V_x V_y \psi(0,0) &= V_x \psi(0,1)= 
 {\rm e}^{2\pi {\rm i} \theta }\psi(1,1)\,.
\end{align}
Equivalently,
\begin{equation}
 V_x^{-1}  V_y^{-1}V_x V_y \psi(0,0) = 
 {\rm e}^{2\pi {\rm i} \theta} \psi(0,0)\,.
\end{equation}
Notice that the charge $2\pi \theta$ plays the role of a central 
charge for the translations on the torus. Moreover, since the final 
result at $(1,1)$ should be independent of the path, we obtain 
$\theta\in\mathds{Z}$. This is electric charge quantization in 
the field theory and magnetic flux quantization in the Landau model.

The quantization of $\theta$ has also implications for the degeneracy 
of the Hamiltonian. In particular, the solutions to the stationary 
Schr\"odinger equation $H\psi_k(x,y)=E\psi_k(x,y)$ which are compatible 
with the boundary conditions (\ref{boundaryconditiononx}) may be written as
\begin{equation}
\label{k-vector-states-1}
 \psi_k(x,y) = \sum_{n\in\mathds{Z}}~\varphi_{nk}(x)~
 {\rm e}^{2\pi {\rm i} \theta (n+\frac{k}{\theta}) y}\,,
 \quad 
 k=0, \dots, \theta-1,
\end{equation}
where 
\begin{equation}
 \varphi_{nk}(x)=f\left(x+n+\frac{k}{\theta}\right),
\end{equation}
with $f$ satisfying a harmonic oscillator equation with 
angular frequency $2\pi \theta$. The energy levels are 
given by $E_\lambda = 2\pi \theta (\lambda+1/2)$ with 
$\lambda \in\mathds{Z}^*$. Note that the wave functions 
are not defined for $\theta=0$, which would correspond to
zero transverse magnetic field.

Observe that the stationary wave functions $\psi_k(x,y)$ 
are also eigenfunctions of the discrete translations 
(\ref{discrete_translations}) in the $y$ direction
\begin{equation}
V_y(l)\psi_k(x, y)=
{\rm e}^{2\pi {\rm i} k l/\theta}
\psi_k(x, y).
\end{equation}
On the other hand, the discrete translations 
(\ref{discrete_translations}) in the $x$ direction 
do not leave $\psi_k(x, y)$ invariant but, defining 
$V_\theta\equiv V_x(1)$, they act as
\begin{equation}\label{V_theta}
V_\theta\psi_k(x, y)=\psi_{k+1}(x, y).
\end{equation}
As pointed out before, one can see that the rest of translations
generated by the operators in (\ref{trans_v}) do not preserve the 
domain of $H$ and map $\psi_k(x, y)$ to a wave function that, in 
general, does not satisfy the boundary conditions 
(\ref{boundaryconditiononx}).

In the following section, we will be interested in the 
ground state solutions, $\lambda=0$, where 
$\varphi_{nk}(x)\sim {\rm e}^{-\pi\theta(x+n+\frac{k}{\theta})^2}$. 
Therefore, the ground state wave functions are given by
\begin{equation}\label{k-ground-state-1}
    \psi_k(x,y)=\mathcal{N}\sum_{n\in\mathds{Z}}
    {\rm e}^{-\pi\theta(x+n+\frac{k}{\theta})^2}
    {\rm e}^{2\pi {\rm i}\theta(n+\frac{k}{\theta})y}\,,
\end{equation}
where $\mathcal{N}$ is a normalization constant. 
The wave function (\ref{k-ground-state-1}) can be 
rewritten using the Jacobi $\vartheta$ function,
$$\vartheta_3(z\,|\tau)=\sum_{n\in\mathds{Z}}
{\rm e}^{{\rm i}\pi \tau n^2+2\pi{\rm i} z n}\,,$$
in the form
$$\psi_k(x, y)=\mathcal{N} {\rm e}^{2\pi{\rm i}k y-\pi\theta(x+k/\theta)^2}
\vartheta_3(\theta({\rm i}x+y+{\rm i}k/\theta)\,|\, {\rm i}\theta)\,.$$

\section{Entanglement Entropy}\label{sec:entanglemententropy}

In this section, we study the entanglement between
the charged particle and the electromagnetic field in the
ground state of the theory described by the Lagrangian
(\ref{lagrangian-ym}). According to the analysis performed
in section \ref{sec:ym}, this is equivalent to measuring the
entanglement between the two degrees of freedom, $x$ and $y$,
of the Landau model on a torus defined by the Hamiltonian
(\ref{Hamiltonian-Landau-on-Torus-1}). We shall compute the
entanglement entropy in the ground state of the latter system.
As we have seen in the previous section, the ground state
is degenerate. We shall then analyse the entanglement entropy
in the degeneracy subspace $\{|\psi_k\rangle\}_{k=0}^{\theta-1}$ 
where the vectors $|\psi_k\rangle$ are such that
$\psi_k(x,y)\equiv \langle x,y | \psi_k\rangle$ with $\{ |x,y\rangle\}$ being
the coordinate basis and $\psi_k(x,y)$ being the ground state wave function
given in (\ref{k-ground-state-1}).

In order to define the entanglement entropy,
we need the associated density matrix
$\rho_k=|\psi_k\rangle\langle \psi_k|$,
whose entries in the coordinate basis are
\begin{equation}\label{density-matrix}
  \rho_k(x,y;x',y')=\overline{\psi_k(x,y)}\psi_k(x',y')\,.
\end{equation}
Now we have to trace out one of the degrees of freedom, say
$y$. This is equivalent to tracing out the degrees of freedom
of the particle in the gauge field theory.
Then we obtain the reduced density matrix $\varrho_k$
associated with the gauge field $A_\mu$,
\begin{equation}\label{reduced-density-mat-1}
 \varrho_k(x,x') = \int_0^1 {\rm d}y~ \rho_k(x,y;x',y)\,.
\end{equation}
Finally, the (von Neumann) entanglement entropy is defined as
\begin{equation}\label{def-ent-ent}
  S_k=-{\rm Tr}(\varrho_k \log \varrho_k)\,.
\end{equation}
If we had chosen to trace out the degree of freedom
$x$ that corresponds to the gauge field, the reduced density
matrix would be associated with the charged particle.
Nevertheless, the resulting entanglement entropy does not
depend on which reduced density matrix we consider.

Inserting the explicit expression of the wave function $\psi_k(x,y)$,
see (\ref{k-vector-states-1}) and (\ref{k-ground-state-1}), we find
that the reduced density matrix (\ref{reduced-density-mat-1}) is of
the form
\begin{equation}\label{reduced-density-mat-2}
  \varrho_k(x,x')=\sum_{n\in\mathds{Z}} \varphi_{nk}(x)\varphi_{nk}(x')
  =\mathcal{N}^{\,2}\sum_{n\in\mathds{Z}}
         {\rm e}^{-\pi\theta\left(x+n+k/\theta\right)^2}
    {\rm e}^{-\pi\theta\left(x'+n+k/\theta\right)^2},
\end{equation}
or, in terms of the $\vartheta$ function,
$$\varrho_k(x, x')=\mathcal{N}^{\, 2}
{\rm e}^{-\pi\theta(x^2+x'^2+2(k/\theta)^2)-2\pi k(x+x')}
\vartheta_3({\rm i}\theta(x+x'+2k/\theta)\,|\,{\rm i}2\theta)\,.$$
Now the direct way to obtain the entanglement entropy
would be to compute the eigenvalues of $\varrho_k$ and
then plug them in (\ref{def-ent-ent}). However,
this is in principle a difficult task that we shall
bypass approximating $\varrho_k$ in two different ways.

\begin{figure}
\centering
\subfloat[\centering]{
  \includegraphics[width=0.65\columnwidth]{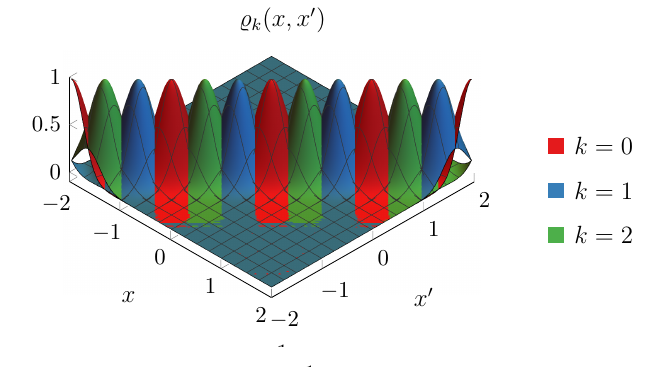}}
  
\subfloat[\centering]{
  \hspace*{-1.2cm}                                                           
  \includegraphics[width=0.8\textwidth]{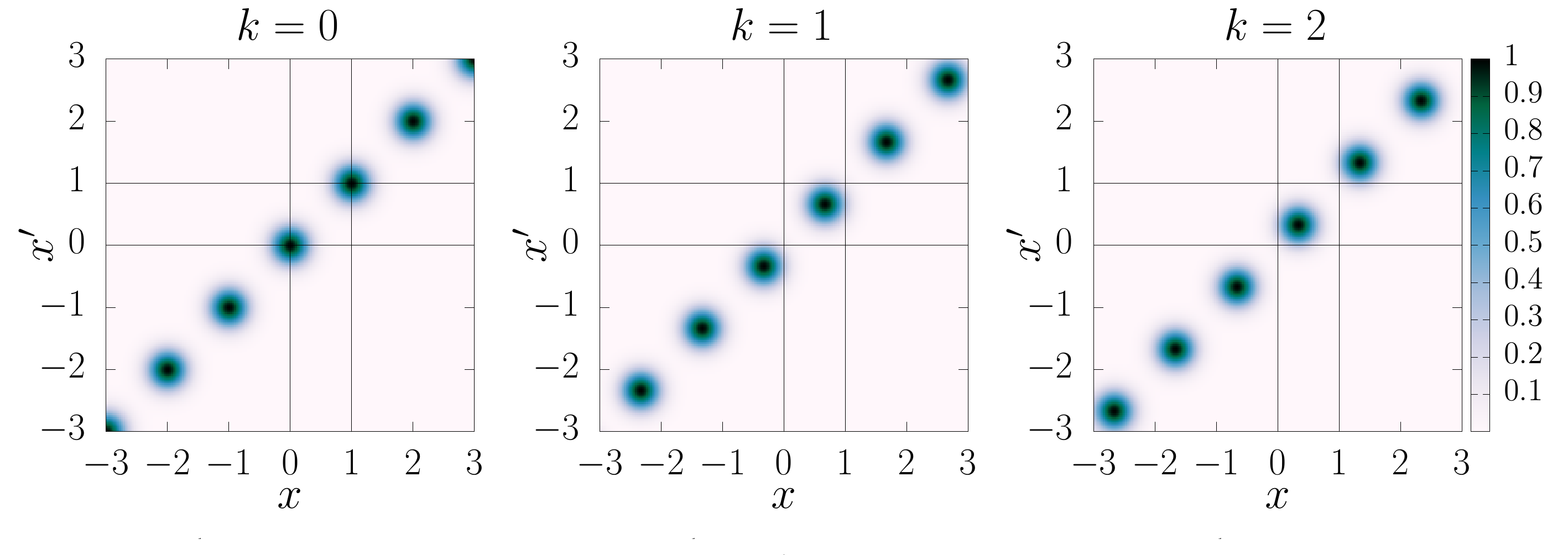}}
\caption{In (a), we represent the reduced density matrix $\varrho_k(x,x')$
  obtained in (\ref{reduced-density-mat-2}) for $k=0, 1, 2$ with $\theta=3$. 
  In (b), we represent the distribution of these three reduced density matrices. 
  Note that, since the particle moves on a torus, the coordinates $x,x'$ are restricted 
  to the interval $[0,1)$, as the lines in (b) indicate.}
\label{fig:f6}
\end{figure}

\begin{figure}
\centering
\subfloat[\centering]{
  \includegraphics[width=0.8\textwidth]{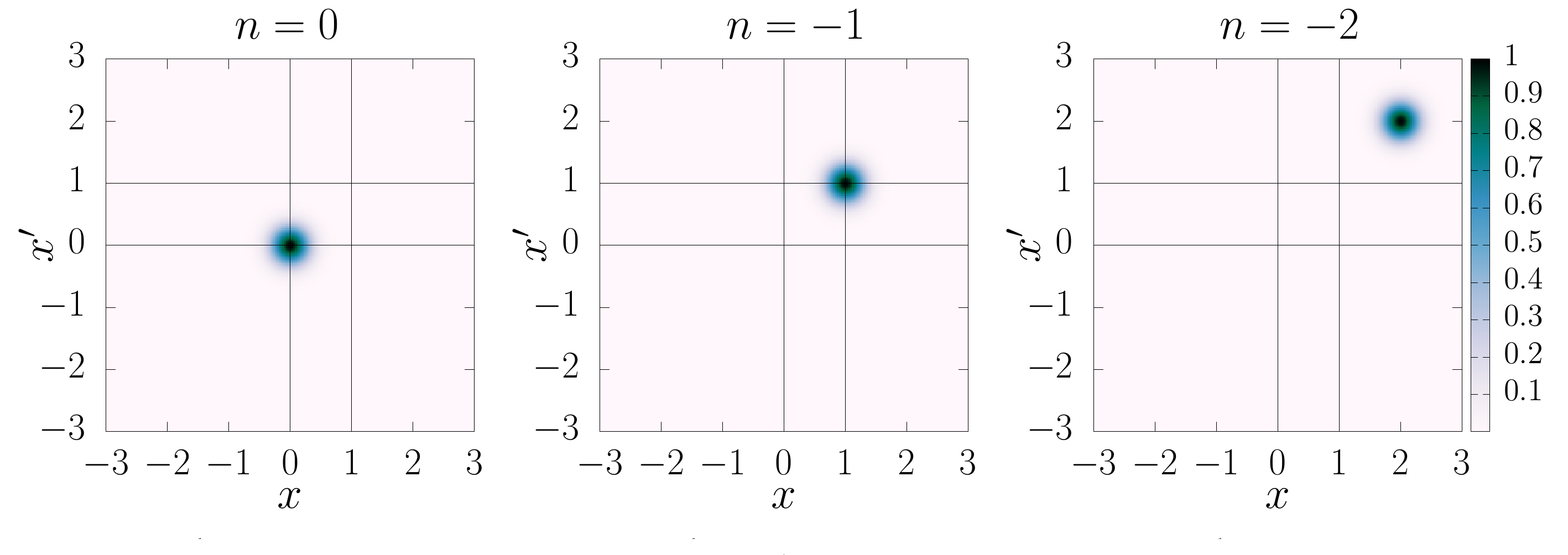}}
  
\subfloat[\centering]{
  \includegraphics[width=0.8\textwidth]{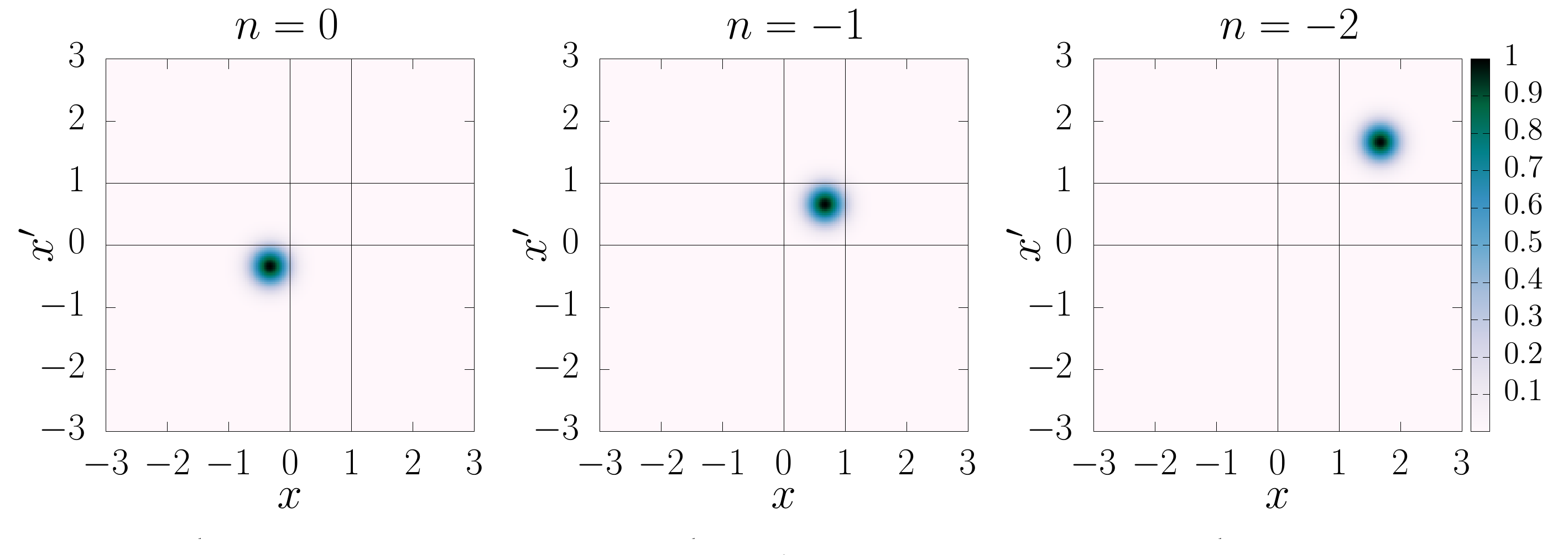}}

\subfloat[\centering]{
  \includegraphics[width=0.8\textwidth]{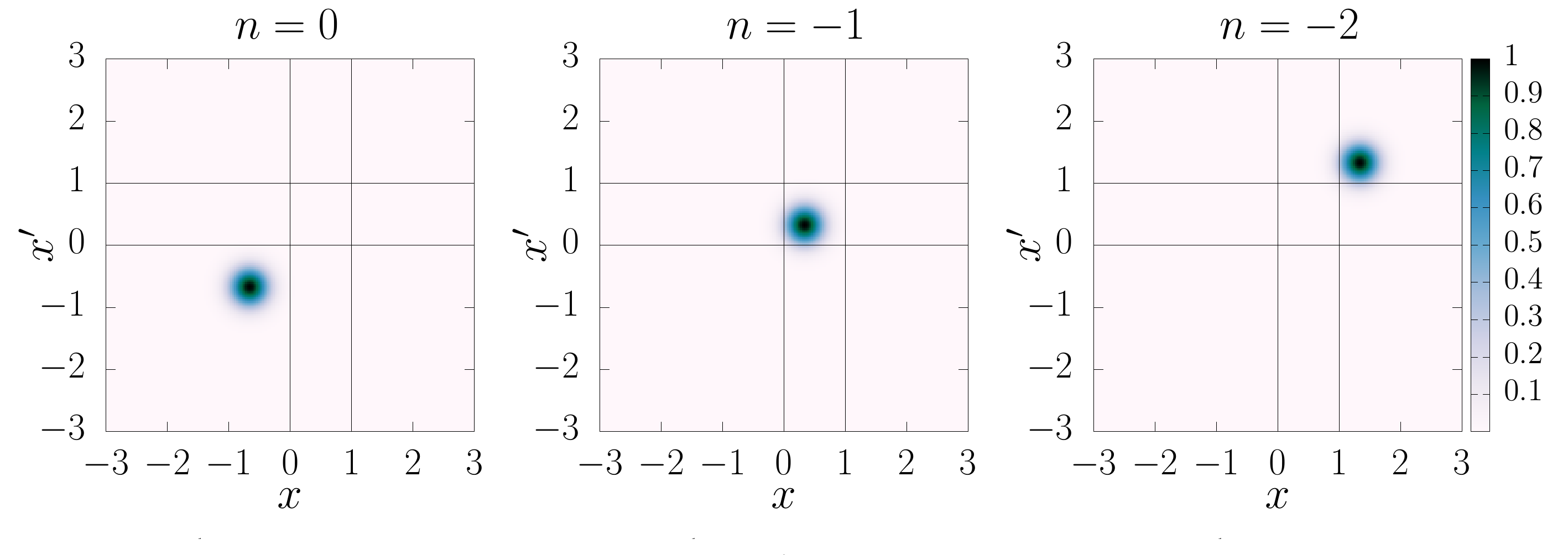}}
\caption{Contribution of the modes
  $n=0, -1, -2$ to the reduced density matrix $\varrho_k(x, x')$,
  see Eq. (\ref{reduced-density-mat-2}), for $\theta=3$ and
  (a) $k=0$, (b) $k=1$, (c) $k=2$.}
\label{fig:f7}

\end{figure}

First, observe that $\varrho_k(x,x')$ is made of peaks localized
along the line $x'=x$, as Fig. \ref{fig:f6} (a) illustrates for
$\theta=3$ and the three possible values for $k$. In Fig.
\ref{fig:f6} (b) we represent separately $\varrho_k(x,x')$ for
each value of $k$ delimiting the interval $[0, 1)$, which
is the domain where the variables $x$, $x'$ are defined.
Observe that as $k$ grows the peaks of $\varrho_k(x,x')$
move down along the line $x'=x$. Each peak of $\varrho_k(x,x')$
comes from one of the modes in the sum (\ref{reduced-density-mat-2}).
Therefore, only the modes that correspond to a peak inside the square
$[0,1)\times[0,1)$ contribute to $\varrho_k(x,x').$ For example,
in Fig. \ref{fig:f7} we can see that for $\theta=3$ the only modes that
contribute are $n=-2, -1, 0$.

In fact, for any value of $k$ and $\theta$, one can see
that only two $n$ modes are significant in the interval
$x\in [0,1)$ so that we can neglect the rest of them in
the calculation of the reduced density matrix. Thus, we
can treat the latter as that of a two-level system. 

For $k/\theta<1/2$, only the peaks corresponding to the modes
$n=0$ and $n=-1$ are relevant in the interval $[0, 1)$.
Therefore, the reduced density matrix can be approximated
as
\begin{equation*}
  \varrho_k(x,x')\approx \varphi_{0k}(x)\varphi_{0k}(x')+
                         \varphi_{-1k}(x)\varphi_{-1k}(x')\,.
\end{equation*}

If $k/\theta>1/2$, then the non-neglectable peaks
in the interval $[0, 1)$ correspond to the modes
$n=-1$ and $n=-2$ and
\begin{equation*}
   \varrho_k(x,x')\approx \varphi_{-1k}(x)\varphi_{-1k}(x')+
                         \varphi_{-2k}(x)\varphi_{-2k}(x')\,.
\end{equation*}

In the case $k/\theta=1/2$, only the peak with $n=-1$, that
is at $x=1/2$, gives a significant contribution.
Thus
\begin{equation*}
   \varrho_k(x,x')\approx \varphi_{-1k}(x)\varphi_{-1k}(x')\,.
\end{equation*}

These approximations may be written in terms of normalised 
functions $u_{nk}(x)$. To do so, we define $p_{nk}$,
$$p_{nk}=\int_0^1 {\rm d} x|\varphi_{nk}(x)|^2=
\frac{{\rm erf}\left(\sqrt{2\pi\theta}
  \left(k/\theta+n+1\right)\right)-
       {\rm erf}\left(\sqrt{2\pi\theta}
       \left(k/\theta+n\right)\right)}{2\sqrt{2\theta}}\,,$$
where ${\rm erf}(z)$ denotes the error function. In terms 
of $p_{nk}$ we can write the approximations as:

\begin{itemize}
\item For $k/\theta < 1/2$,
\begin{equation*}
\varrho_k(x,x')\approx \lambda_{0k} u_{0k}(x)u_{0k}(x') + \lambda_{-1k} u_{-1k}(x)u_{-1k}(x')
\end{equation*}
where
\begin{equation*}
\lambda_{nk}=\frac{p_{nk}}{p_{0k}+p_{-1k}} \quad\mbox{and}\quad u_{nk}(x)=\frac{1}{\sqrt{\lambda_{nk}}}\varphi_{nk}(x)\,.
\end{equation*}
\item For $k/\theta > 1/2$,
\begin{equation*}
\varrho_k(x,x')\approx \lambda'_{-1k} u'_{-1k}(x)u'_{-1k}(x') + \lambda'_{-2k} u'_{-2k}(x)u'_{-2k}(x')
\end{equation*}
where
\begin{equation*}
\lambda'_{nk}=\frac{p_{nk}}{p_{-1k}+p_{-2k}} \quad\mbox{and}\quad u'_{nk}(x)=\frac{1}{\sqrt{\lambda'_{nk}}}\varphi_{nk}(x)\,.
\end{equation*}
\item For $k/\theta = 1/2$,
\begin{equation*}
\varrho_k(x,x')\approx u_{-1k}(x)u_{-1k}(x')\,.
\end{equation*}
\end{itemize}

Note that it follows from the definitions of $\lambda_{kn}$ and
$\lambda'_{kn}$ that $\lambda_{-1k}=1-\lambda_{0k}$ and
$\lambda'_{-2k}=1-\lambda'_{-1k}$. Therefore, the entanglement entropy
can be expressed as

\begin{equation}\label{approx-ent}
  S_k \approx \begin{cases}
    -\lambda_{0k}\log\lambda_{0k}-(1-\lambda_{0k})\log(1-\lambda_{0k})\,, & \mbox{if } k/\theta<1/2\,,\\
    -\lambda_{-1k}\log\lambda_{-1k}-(1-\lambda_{-1k})\log(1-\lambda_{-1k})\,, & \mbox{if } k/\theta>1/2\,,\\
    0\,, & \mbox{if } k/\theta=1/2\,.
  \end{cases}
\end{equation}
The latter case only happens when $\theta$ is an even
number. 

Using the identity between the error function 
and the confluent hypergeometric function
of the first kind $M(a, b, z)$ (see, e.g., Eq.
13.6.7 in \cite{NIST}),
\begin{equation*}
  {\rm erf}(z)=\frac{2z}{\sqrt{\pi}}
  M\left(\frac{1}{2}, \frac{3}{2}, -z^2\right),
\end{equation*}
we have
\begin{equation*}
  p_{nk}=\left(\chi_k+n+1\right)
  \mathrm{M}\left(-2\pi\theta\left(\chi_k+1+n\right)^2\right)-
  \left(\chi_k+n\right)\mathrm{M}\left(-2\pi\theta
  \left(\chi_k+n\right)^2\right),
\end{equation*}
where we have introduced the notation ${\rm M}(z)\equiv M(1/2, 3/2, z)$
and $\chi_k=k/\theta$. Hence we find that
\begin{equation*}
  \lambda_{0k}=
  \frac{(\chi_k+1){\rm M}\left(-2\pi\theta(\chi_k+1)^2\right)-
    \chi_k{\rm M}\left(-2\pi\theta\chi_k^2\right)}
       {(\chi_k+1){\rm M}\left(-2\pi\theta(\chi_k+1)^2\right)-(\chi_k-1)
         {\rm M}\left(-2\pi\theta(\chi_k-1)^2\right)}\,,
\end{equation*}
and
\begin{equation*}
  \lambda'_{-1k}=
  \frac{\chi_k{\rm M}\left(-2\pi\theta\chi_k^2\right)-
    (\chi_k-1){\rm M}\left(-2\pi\theta(\chi_k-1)^2\right)}
       {\chi_k{\rm M}\left(-2\pi\theta(\chi_k)^2\right)-(\chi_k-2)
         {\rm M}\left(-2\pi\theta(\chi_k-2)^2\right)}\,.
\end{equation*}
From these expressions it is clear that the entanglement
entropies for $k/\theta<1/2$ and for $k/\theta>1/2$ are
related by the transformation $k/\theta\mapsto 1-k/\theta.$

Let us check numerically the accuracy of the above results.
This is done by expanding the functions $\varphi_{nk}(x)$ in
Fourier modes,
\begin{equation*}
  \varphi_{nk}(x)=\sum_{p\in\mathds{Z}}
  \tilde{\varphi}_{nk}(p){\rm e}^{2\pi{\rm i}px}\,.
\end{equation*}
In the basis of Fourier modes the entries of the
reduced density matrix (\ref{reduced-density-mat-1})
are given by
\begin{equation}\label{red-mat-fourier}
  \tilde{\varrho}_k(p,p')=\int {\rm d}x {\rm d}x' \varrho_k(x,x')
  {\rm e}^{2\pi{\rm i}px}{\rm e}^{-2\pi {\rm i} p'x'},\quad p, p'\in \mathds{Z}\,.
\end{equation}
In order to compute numerically the entanglement entropy
we truncate the matrix $(\tilde{\varrho_k}(p,p'))$ restricting
the indices $p,p'\in\mathds{Z}$ to the interval $-N\leq p, p'\leq N$.
Then we calculate the eigenvalues of this sub-matrix and
we plug them in the expression of the entanglement entropy
(\ref{def-ent-ent}).

\begin{figure}
\centering
\includegraphics[clip,width=0.7\columnwidth]{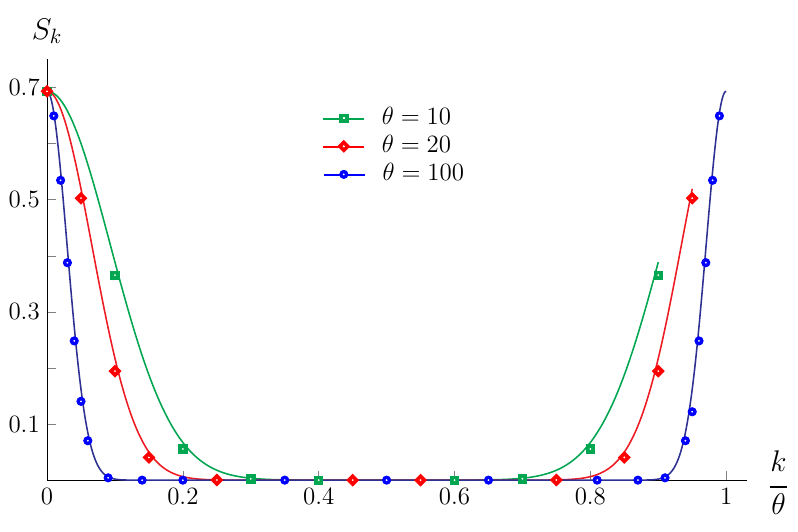}
\caption{Entanglement entropy as a function of the ratio $k/\theta$
  for several fixed values of $\theta$ and varying $k$. The solid lines
  represent the analytical approximation obtained in Eq. (\ref{approx-ent}). The
  dots have been obtained numerically from the matrix (\ref{red-mat-fourier}),
  taking as cut-off $N=100$ .}
\label{fig:f3}

\end{figure}

The value obtained numerically for the entanglement entropy
should converge to that predicted by the expression (\ref{approx-ent})
as we increase the cut-off $N$. In Fig. \ref{fig:f3} we compare the results for
a given $\theta$ and $k$ varying from 0 to $\theta-1$. As we can see the
results agree for $N$ large enough. Notice also that the entanglement
entropy varies with $k$. This means that there is an ambiguity associated
to the entanglement entropy of the ground state.

\begin{figure}[h]
\centering
\includegraphics[clip,width=0.8\columnwidth]{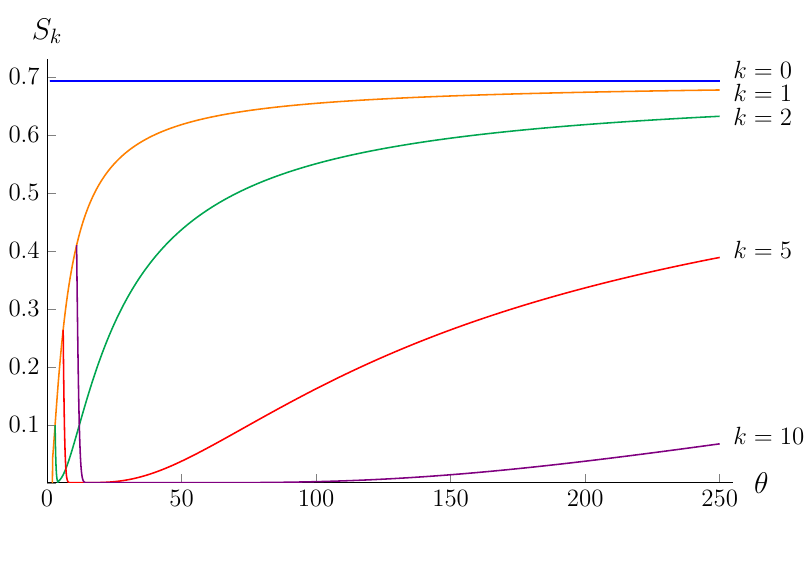}
  \caption{Entanglement entropy as a function of $\theta$ for some fixed values 
   of $k$ using the analytical approximation found in (\ref{approx-ent}).}
\label{fig:f5}
\end{figure}

It is also interesting to analyse how the entanglement
entropy behaves as a function of $\theta$ for fixed $k$.
Recall that $\theta$ is proportional to the electric charge of the
particle (we have set $q=-2\pi\theta$). In Fig. \ref{fig:f5}, we plot $S_k$ in terms of $\theta$
for several fixed values of $k$ using the analytical approximation
(\ref{approx-ent}). The initial point of the curve for each $k$ corresponds to
$\theta=k+1$. Observe that, due to the symmetry $k/\theta\mapsto 1-k/\theta$,
the initial points of all the curves with $k>1$ also belong to the
curve for $k=1$. As $\theta$ increases, $S_k$ decreases
until $\theta=2k$, where it vanishes. From this
point, $S_k$ increases tending to $\log 2$ when $\theta\to \infty$
(infinite transverse magnetic flux in the associated Landau model).
Note that as $k$ is larger, the entropy saturates more slowly to the
asymptotic value $\log 2$. We can conclude that there is an upper
bound for the entanglement entropy in the states $|\psi_k\rangle$, 
which is exactly that of a maximally entangled two-level quantum system. 

Another interesting case is $\theta=0$. It corresponds
to a particle with zero electric charge (zero transverse
magnetic flux in the Landau model). For $\theta=0$, the analytical
approximation (\ref{approx-ent}) is not well defined. Nevertheless, since the particle
and the gauge field are decoupled, the degrees of
freedom of the Landau model, $x$ and $y$, are separable; that is,
the wave function of the ground state, that now is not degenerate,
can be factorized in the form $\psi(x, y)=X(x)Y(y)$. This implies that
the entanglement entropy is zero for $\theta=0$. 

\subsection{Entanglement Entropy for a $V_\theta$-invariant State}

We have just seen that the entanglement entropy changes inside the 
degeneracy space of the ground state; that is, 
according to Eq. (\ref{V_theta}), it varies 
under the discrete translations in the $x$ 
direction of the torus defined by $V_\theta$. 
However, there is a state which is invariant under 
$V_\theta$: the linear combination of the states 
$\{|\psi_k\rangle\}_{k=0}^{\theta-1}$ 
that span the ground state degeneracy subspace
\begin{equation}
  |\psi_\theta\rangle=
  \frac{1}{\sqrt{\theta}}
  \sum_{k=0}^{\theta-1}|\psi_k\rangle.
\end{equation}
In the coordinate basis, its density matrix 
$\rho_\theta=|\psi_\theta\rangle\langle \psi_\theta|$ 
reads
\begin{equation}\label{democratic_rho}
 \rho_\theta(x,y;x',y') = \frac{1}{\theta}
 \sum_{k, k'=0}^{\theta-1}
 \overline{\psi_k(x,y)}\psi_{k'}(x', y').
\end{equation}

In order to calculate the entanglement entropy 
of $|\psi_\theta\rangle$, we can take the partial 
trace in (\ref{democratic_rho}) with respect to 
either $x$ or $y$. It is more convenient to 
take it with respect to the former. The corresponding
reduced density matrix,
\begin{equation*}
\sigma_\theta(y, y')=
\int_0^1 {\rm d}x ~
\rho_\theta(x, y; x, y'),
\end{equation*}
can be expressed in the form
\begin{equation*}
\sigma_\theta(y, y')= \sum_{p, p'\in \mathds{Z}} 
\tilde{\sigma}_\theta(p, p')
{\rm e}^{-2\pi {\rm i} p y}{\rm e}^{2\pi{\rm i}p'y'},
\end{equation*} 
with
\begin{equation}\label{red_dens_mat_inv_state_y}
\tilde{\sigma}_\theta(p, p')=
\mathcal{N}^{\,2}\frac{{\rm e}^{-\pi(p-p')^2/(2\theta)}}
{2\theta\sqrt{2\theta}}
\left({\rm erf}\left(\sqrt{\frac{\pi}{2\theta}}
(2\theta+p+p')\right)
-{\rm erf}\left(\sqrt{\frac{\pi}{2\theta}}
(p+p')\right)\right).
\end{equation}

\begin{figure}[h]
\centering
\includegraphics[clip,width=0.8\columnwidth]{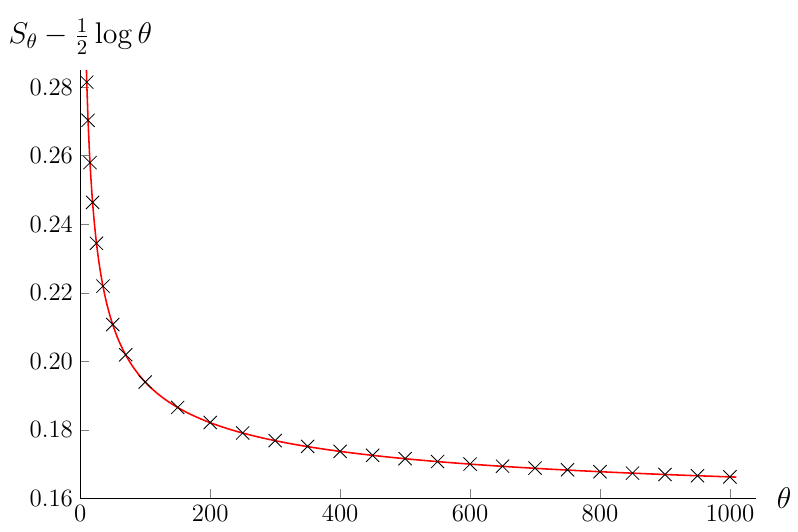}
  \caption{Entanglement entropy of the $V_\theta$-invariant 
  state (\ref{democratic_rho}) substracting the term $1/2\log\theta$ 
  as a function of $\theta$. The crosses correspond to the values 
  obtained for $S_\theta$ by diagonalizing numerically 
  the reduced density matrix (\ref{red_dens_mat_inv_state_y}), 
  restricting the indices to $-1100\leq p, p'\leq 1100$.
  The solid line is the curve $\alpha+\beta/\sqrt{\theta}$ fitted to the numerical points
  in the interval $\theta\in[500, 1000]$; we obtain $\alpha=0.15343$
  and $\beta=0.404832$.
  }
\label{fig:ent_inv_state}
\end{figure}

Observe that the elements $\tilde{\sigma}_\theta(p, p')$, $p,p'\in\mathds{Z}$, 
form an infinite matrix which represents $\sigma_{\theta}$ 
in the momentum space. As we have done before, we can calculate
the entanglement entropy $S_\theta$ of $|\psi_\theta\rangle$ 
from the spectrum of this matrix truncated for $|p|$, $|p'|$ 
large enough. In Fig. \ref{fig:ent_inv_state}, we have computed numerically 
$S_\theta$ varying $\theta$ using this method. From this plot, 
we can conclude that
\begin{equation}\label{ent_inv_state}
S_\theta\sim 
\frac{1}{2}\log\theta
+\alpha
+\frac{\beta}{\sqrt{\theta}},
\quad \mbox{for}
\quad \theta\gg 1.
\end{equation}
The value of the coefficients $\alpha$, $\beta$, see the caption of
Fig. \ref{fig:ent_inv_state}, can be determined from the fit of (\ref{ent_inv_state})
to the numerical data. Note in this figure that the fitted curve 
$\alpha+\beta/\sqrt{\theta}$ is very close to the numerical data. 
This is specially remarkable if we take into account that the fit 
was performed for values of $\theta$ between 500 and 1000 and 
then the curve is plotted from $\theta=10$.

Therefore, contrary to the states $|\psi_k\rangle$,
eigenfunctions of the translations $V_y(l)$ in the 
$y$ direction of the torus, whose entanglement entropy 
tends to $\log 2$ in the limit $\theta\to\infty$ 
(see Fig. \ref{fig:f5}), the entanglement entropy of 
the state $|\psi_\theta\rangle$, invariant under the 
translations $V_x(l)$ in the $x$ direction, diverges 
logarithmically with the transverse magnetic flux/charge 
of the particle.

It is worth commenting that another state 
invariant under $V_\theta$ is the mixed state described 
by the density matrix 
$$\rho_{\rm M}=\frac{1}{\theta}
\sum_{k=0}^{\theta-1} 
|\psi_k\rangle\langle\psi_k |,$$
or, in the coordinate basis,
\begin{equation}\label{inv_mixed_state}
\rho_{\rm M}(x, y; x', y')=
\frac{1}{\theta}
\sum_{k=0}^{\theta-1}
\rho_k(x, y; x', y').
\end{equation}
It corresponds to the equiprobable classical ensemble 
of the states $\{|\psi_k\rangle\}_{k=0}^{\theta-1}$. 
It is easy to see that the partial traces of $\rho_{\rm M}$ 
and $\rho_\theta$ with respect to $y$,
$$\varrho_{\rm M}(x, x')=
\int_0^1{\rm d}y ~
\rho_{\rm M}(x, y; x', y), \quad
\varrho_\theta(x, x')=
\int_0^1{\rm d}y ~ 
\rho_\theta(x, y; x', y),$$
lead to the same reduced density matrix
$$\varrho_\theta(x, x')=
\varrho_{\rm M}(x, x')=
\frac{1}{\theta}\sum_{k=0}^{\theta-1}
\varrho_k(x, x').$$
This means that, from the perspective 
of the field theory problem, the gauge field
cannot distinguish if the whole system 
is in the linear combination or in the statistical 
ensemble of the states $\{|\psi_k\rangle\}_{k=0}^{\theta-1}$. 
The same is not true for the particle. The partial trace 
of (\ref{inv_mixed_state}) with respect to $x$,
$$
\sigma_{\rm M}(y, y')=
\int_0^1
{\rm d} x ~ 
\rho_{\rm M}(x, y; x, y'),$$
can be written in the momentum space as
$$\tilde{\sigma}_{\rm M}(p, p')=
\delta_{p, p'}^{{\rm mod}\,\theta}
\tilde{\sigma}_{\theta}(p, p'),$$
where $\delta_{p, p'}^{{\rm mod}\,\theta}$
is 1 if $p=p'\,({\rm mod}\,\theta)$ and 0 otherwise. 
One can numerically check that the entropies obtained from 
$\sigma_{\rm M}$ and $\sigma_{\theta}$ are indeed different.

\section{Conclusions}\label{sec:conclusions}

In this work, we studied the ground state entanglement
entropy between an electromagnetic field and a charged 
non-relativistic particle on a space-time cylinder. In 
order to compute this entropy, we resorted to the fact 
that a Yang-Mills field theory defined on a space-time
cylinder can be mapped to the problem of a free quantum 
particle moving on the gauge group manifold. In our case, 
we considered an electromagnetic field, for which the 
gauge group is $U(1)$, and therefore the corresponding 
manifold is the unit circle. Since here the gauge field 
is coupled to a non-relativistic particle, the 
associated quantum mechanical problem is a particle moving 
on a torus with a transverse magnetic field: the Landau 
model on a torus. The two degrees of freedom of the particle 
on the torus correspond respectively to the gauge field and 
to the non-relativistic particle in the field theory problem.

Therefore, the computation of the entanglement entropy between 
the electromagnetic field and the non-relativistic particle
was reduced to taking the partial trace of one of the
degrees of freedom in the wave function of the particle
on the torus and computing the entropy from the corresponding
reduced density matrix. Since the ground state of the Landau model
is degenerate, we analysed the entanglement entropy
in the degeneracy subspace. We performed this analysis
treating the reduced density matrix of the states that generate
this subspace as that of a two-level system. We obtained an
approximate analytical expression for their entanglement entropy
which was checked numerically. In particular, we  found
that, when the electromagnetic field and the particle are
decoupled, the entanglement entropy is zero while, when the
particle's charge goes to infinity, the entanglement
entropy tends to $\log 2$. 

The symmetry behind the degeneracy of the ground
state is the group of translations $\mathds{Z}_\theta$ in the $y$ 
direction of the torus. The translational symmetry 
is anomalously broken due to the boundary conditions of the Landau Hamiltonian
to the discrete subgroup $\mathds{Z}_\theta\times \mathds{Z}_\theta$, 
where $\theta$ is the electric charge of the particle in the field 
theory/the transverse magnetic flux in the Landau model, which is 
quantized. We also studied the entanglement of the state
invariant under the $\mathds{Z}_\theta$ translations in the $x$ direction 
of the torus. This state can be constructed from the equiprobable 
linear combination of the states that generate the ground state
degeneracy subspace. In this case, the entanglement entropy does not saturate
to a finite value when the particle's charge goes to infinity, but 
it scales logarithmically with the charge.

The natural continuation of this work is to take a 
non-Abelian Yang-Mills theory instead of an Abelian one
and study how the results obtained here generalise to the
$SU(N)$ gauge group. In particular, the YM theory would 
be mapped to a particle moving along a different gauge 
group manifold. For example, for $SU(2)$ we would have 
a particle moving along $S^3$. Solving its dynamics would 
then mean working with a non-trivial set of Wong's equations 
\cite{wong}. Another interesting aspect to analyse is the 
evolution of the entanglement between the matter and the 
gauge sectors after a quantum quench \cite{Kumar1,Kumar2}. 
This could be done by preparing the system in the ground state 
in which the gauge field and the particle are decoupled and then suddenly 
turning on the interaction term, for example. Indeed, the 
non-equilibrium dynamics of a 1+1 dimensional $U(1)$ gauge 
theory coupled either to fermions \cite{Brenes} or to bosons 
\cite{Chanda} has recently been investigated and it was observed that
the system may not thermalize. We plan to tackle these problems 
in the future.

\section*{Acknowledgements}

We thank A. Melikyan,  A. Pinzul and D. Trancanelli for valuable comments on the manuscript.
FA acknowlegdes financial support from the Brazilian ministries MEC and MCTIC.
MT acknowledges the support of the Conselho Nacional de Desenvolvimento
Cientifico e Tecnologico (CNPq). We thank the anonymous referee of JHEP for 
the interesting comments 
and questions which have allowed us to improve this work.

\appendix
\section{Green's function}
\label{appendix}

We want to solve $\partial^2_s G(s,s')=\delta(s-s')$
in the domain $s, s'\in \left[0,2\pi R\right)$ with periodic 
boundary conditions $G(s+2\pi R, s')=G(s, s'+2\pi R)=G(s, s')$. 
Since the system is translationally invariant, $G(s, s')\equiv G(s-s')$.
Therefore, we can rewrite the problem in terms of the 
variable $\texttt{s}=s-s'$ as $\partial^2_\texttt{s} 
G(\texttt{s})=\delta(\texttt{s})$ with boundary condition 
$G(\texttt{s})=G(\texttt{s}+2\pi R)$. In the end, we will 
just need to replace $\texttt{s}$ by $s-s'$.

In order to solve $\partial^2_\texttt{s} G(\texttt{s})=\delta(\texttt{s})$, 
we express $G(\texttt{s})$ and $\delta(\texttt{s})$ in terms of 
their Fourier series,
\begin{equation}
	G(\texttt{s})=a_0 
	+ \sum_{n\neq 0} a_n {\rm e}^{{\rm i} n\texttt{s}/R}, 
	\quad\mbox{and}\quad
	\delta(\texttt{s})=\frac{1}{2\pi R}
	+\frac{1}{2\pi R}\sum_{n\neq 0}{\rm e}^{{\rm i}n\texttt{s}/R}.
\end{equation}
After plugging them in the differential equation, we obtain
\begin{equation}\label{green_fourier}
	\sum_{n\neq 0}\left(\frac{{\rm i}n}{R}\right)^2 a_n 
	{\rm e}^{{\rm i}n\texttt{s}/R} 
	\buildrel!\over=   
	\frac{1}{2\pi R}+\frac{1}{2\pi R}\sum_{n\neq 0}
	{\rm e}^{{\rm i}n\texttt{s}/R}.
\end{equation}
Note that the presence of a zero mode on the right-hand 
side is problematic. To circumvent this issue, we remove 
it for now. We will discuss the validity of 
this later. Going back to the differential equation, we 
should now solve
\begin{equation}
	\sum_{n\neq 0}\left(\frac{{\rm i}n}{R}\right)^2
	a_n {\rm e}^{{\rm i}n\texttt{s}/R}
	=\frac{1}{2\pi R}\sum_{n\neq 0}
	{\rm e}^{{\rm i}n\texttt{s}/R},
\end{equation}
which gives $a_n=-\dfrac{R}{2\pi n^2}$. Therefore,
\begin{equation}
	G(\texttt{s})=
	-\sum_{n\neq 0} 
	\frac{R}{2\pi n^2}{\rm e}^{{\rm i}n\texttt{s}/R}.
\end{equation}

Now the problem boils down to computing this infinite sum. 
To do so, we note that
\begin{eqnarray}
	G(\texttt{s})= 
	- \sum_{n=-\infty}^{-1} \frac{R}{2\pi n^2} {\rm e}^{{\rm i}n\texttt{s}/R} 
	-\sum_{n=1}^{\infty}\frac{R}{2\pi n^2} {\rm e}^{{\rm i}n\texttt{s}/R}
	&=&-\frac{R}{2\pi}\sum_{n=1}^{\infty}\left(\frac{{\rm e}^{-{\rm i}n\texttt{s}/R}}{n^2}
	+\frac{{\rm e}^{{\rm i}n\texttt{s}/R}}{n^2}\right)\nonumber\\
	&=&-\frac{R}{2\pi}\left(\mbox{Li}_2({\rm e}^{-{\rm i}\texttt{s}/R})
	+\mbox{Li}_2({\rm e}^{{\rm i}\texttt{s}/R})\right),
\end{eqnarray}
where $\mbox{Li}_2(z)$ is the dilogarithm function. 
Using the identity (see, e.g., Eq. 25.12.4 in \cite{NIST}),
\begin{equation}
	\mbox{Li}_2(z)+\mbox{Li}_2\left(\frac{1}{z}\right)
	=-\frac{\pi^2}{6}-\frac{1}{2}\left(\ln(-z)\right)^2,
\end{equation}
which for the case $z={\rm e}^{{\rm i}\phi}$, $\phi\in[-2\pi, 2\pi]$,
reads
$$\mbox{Li}_2({\rm e}^{{\rm i}\phi})
+\mbox{Li}_2({\rm e}^{-{\rm i}\phi})
=-\frac{\pi^2}{6}+\frac{1}{2}(|\phi|-\pi)^2,$$
we finally obtain
\begin{equation}
	G(\texttt{s})=a_0
	-\frac{\texttt{s}^2}{4\pi R}
	+\frac{|\texttt{s}|}{2}
	-\frac{\pi R}{6}. \label{G}
\end{equation}
with $\texttt{s}\in[-2\pi R, 2\pi R]$.
The zero mode $a_0$ is irrelevant and we 
take it to be zero. 

In terms of $s,s'\in \mathds{R}$, we have
\begin{equation}
	G(s,s')=
	-\frac{([s]-[s'])^2}{4\pi R}
	+\frac{|[s]-[s']|}{2}
	-\frac{\pi R}{6},
\end{equation}
satisfying
\begin{equation}
\partial_s^2 G(s,s')=\delta(s-s')-\frac{1}{2\pi R}\,.
\end{equation}

Finally, we mention that removing the zero mode 
of the Dirac delta does not compromise the result. 
In fact, if we act with $\partial_s^2$ on (\ref{A0-sol-w-Green-1}),
we obtain
\begin{eqnarray*}
\partial_s^2 A_0(s)=\int {\rm d}s'~ \partial_s^2 G(s,s') \rho(s') 
&=& \int {\rm d}s'~ \bigg(\delta(s-s')-\frac{1}{2\pi R}\bigg) \rho(s')\\
&=& \rho(s) - \frac{1}{2\pi R} \int {\rm d}s' \rho(s')\,.
\end{eqnarray*}
Since the electric field satisfies periodic boundary 
conditions, the last term vanishes,
\begin{equation}
\int {\rm d}s'~\rho(s')=\int {\rm d}s' ~\partial_{s'} E(s') = 0\,.
\end{equation}
Thus $G(s,s')$ indeed solves the Gauss law constraint.


\end{document}